\documentstyle[12pt,a4,epsf]{article}

\epsfclipon

\def\dfrac#1#2{{\displaystyle#1\over\displaystyle#2}}

\title{The parallax distorsion via a weak microlensing effect}
\author{M.V. Sazhin, V.E. Zharov, T.A. Kalinina\\
Sternberg Astronomical Institute, Moscow 119899, Russia}

\date{
}

\begin{document}

\maketitle

\begin{center}
Abstract
\end{center}

Parallax measurements allow distances to celestial objects to be determined.
Coupled with measurement of their position on the celestial sphere, it gives
a full three-dimensional picture of the location of the objects relative to
the observer. The distortion of the parallax value of a remote source
affected by a weak microlensing is considered. This means that the weak
microlensing leads to distortion of the distance scale. It is shown that the
distortions to appear may change strongly the parallax values in case they
amount to several microseconds of arc. In particular, at this accuracy many
measured values of the parallaxes must be negative.

\section*{Introduction}

Very long baseline interferometry (VLBI) has achieved a  precision level of
position measurements of tens of microseconds of arc \cite{fei99},
\cite{eub98}. The VLBI accuracy may soon achieve several microseconds of arc,
or the fundamental limit of accuracy of the position measurements being
determined by a nonstationary curvature of space-time in our Galaxy
\cite{saz96}, \cite{saz98}.

Besides, creation of space interferometers much exceeding the Earth's
diameter is in sight \cite{and81}, \cite{kar86}, \cite{and86}. A 10 -- 100
times increase in the interferometer baseline is considered to allow a
precision of position measurements of the order of one microsecond of arc
($ \sim 1 \mu $as), or even one hundred nanoseconds of arc ($ \sim
100 $ nas) to be achieved.

In optical ground-based astronomy until the recent decade the position
accuracy has been amounted to $ \sim 0"1$, which was worse by far than
in radioastronomy. A great success of the space project HIPPARCOS lies
in the precision of $ \sim 1\; mas$ having achieved in optical
astronomy for measuring stellar coordinates and parallaxes. The
astronomers hope to develop this success in the space experiments being
planned \cite{gai98}, \cite{sim98}, \cite{dar98}, \cite{faim},
\cite{diva}. A precision of $ \sim 1 \div 10 \; \mu as$ is planned to
be achieved.  High-precision angle measurements allow the stellar
distance scale to be increased from 1 kpc to several tens and hundreds
of kpc.

The inclusion of general relativistic effects has become a necessary part of
the observation reduction procedure when the accuracy of observations is
close to the value of $ \sim 1\; mas$ (of the order of a millisecond of
arc).  The reduction procedure involved gravitational effects induced
by the Sun and planets of the solar system \cite{iers96}. These bodies
induce a nonstationary curvature of space-time in the solar system
making a ray from the celestial source move along a curved trajectory.
Positions, velocities and masses of the Sun and planets are known with
a high degree of accuracy, which provides a possibility of allowing for
the gravitational effects precisely in position measurements.

As the accuracy increases, the astronomers will undoubtely encounter
new phenomena. One of them, the nonstationarity of space-time being due
to the motion of visible stars and invisible bodies in our Galaxy, was
discussed in papers \cite{saz96}, \cite{saz98}. The list of phenomena
which will be important in microarcsecond and sub-microarcsecond
astrometry is discussed in \cite{kop00}.

The nonstationary curvature, induced by the Sun and planets, is a determinate
process. The nonstationary curvature, created by moving stars in our Galaxy,
is a stochastic process. This process is stochastic since distances to most
of  (in particular, invisible) stars as well as their masses are unknown to
the observer. Since these values are unknown, it is impossible to reduce
the observations with the same degree of definiteness as in the solar system.
For the observer a remote source will execute a stochastic motion under the
action of this process about the average position - the true position of the
source in the sky. The value of this oscillation is of the order of
$1 \mu as$, the characteristic time from tens to hundreds of years.

Since the reduction procedure becomes impossible, the value equal to $
\sim 1 \mu as$ was called a fundamental limit of position measurement
accuracy \cite{saz96}, and the effect itself was called a weak
microlensing effect \cite{saz98}.

In the previous papers the weak microlensing effect has been discussed for a
single observer. However, a very important astronomical information is given
simultaneously by two observers (or one observer at different instants) from
different points of space. Such observations allow a celestial source
parallax to be measured and a distance to the source to be determined.
Clearly the nonstationary curvature of space-time in our Galaxy will distort
parallax measurements. This paper is devoted to discussion of the weak
microlensing effect on the parallax measurements.

The parallax measurement is performed as follows: two telescopes spaced at a
distance, called the baseline $B$ (or one telescope in different epochs
realizing the baseline due to the Earth's motion), observe one cosmic object.
Clearly the direction to the object of observation is different for each
telescope. The difference of these directions is called an object parallax
shift $p$ and is determined as follows (provided that at one end of the
baseline the direction to the source is perpendicular to the baseline):
$$
p= \frac{B}{r}
$$

\noindent
where $r$ is the distance to the source. The parallax is a natural unit
of distance measurement in astronomy provided the baseline is the
Earth's orbit equal to one astronomical unit (AU). The parallax equal
to one second of arc ($1"=5 \cdot 10^{-6}$rad) at the baseline of 1 AU
means that the distance to the star being observed is equal to 1 pc
($3 \cdot 10^{18}$cm).  Accordingly, the parallax equal to one
nanosecond of arc ($5 \cdot 10^{-15}$) means that the distance to the
source of light is equal to 1 gigaparsec, which is a good part of the
distance to the horizon of particles of our Universe. The measurement
of distances to extragalactic sources assumes measurements of
ultrasmall angles.  Hence very long baselines and the limiting accuracy
of measurements are necessary for parallax measurements to be realized.

The effects related to a nonstationary curvature of space-time in our Galaxy
will already become important at this level of measurement accuracy. Such
effects, applied to extragalactic astronomy, were considered even 30 years
ago \cite{zel64}, \cite{ref64}, \cite{das65}, see also \cite{bli89}.

As shown below, the nonstationary gravitational field of moving stars in our
Galaxy creates a curvature of space-time sufficient to distort strongly the
values of remote source parallaxes. This effect will have a considerable
impact on parallax measurements of many extragalactic sources and should be
taken into account in future measurements and interpretations.

Variations in the space-time curvature induced by a nonstationary motion of
stars in our Galaxy lead to distortion of the visible position of
extragalactic sources \cite{saz96}, \cite{saz98}, \cite{zda95}. The value of
these distortions is estimated to be about one microsecond of arc for a
considerable number of remote sources. The measurements of positions of these
sources, conducted in considerable time intervals (several tens of years),
may show the presence of such a shift ($\sim 1 \mu as$). This may be
interpreted either as a real shift of the source in space due to a
proper motion or as a visible shift due to the weak microlensing
effect.

Random shifts of the order of one microsecond of arc do not allow the
parallax method to be used any longer for measuring  distances to light
sources exceeding 1 megaparsec. This is about the distance to the Andromeda
galaxy, one of the nearest galaxies to us. The microlensing effect bounds the
distance scale that may be obtained from position measurements.

Measurements of the annual parallax (i.e. conducting measurements in half a
year), or parallax measurements by inteferometry methods, will be more exact
since variations in the Galaxy's nonstationary field will be much less.

However it should be noted that phenomena related to the nonstationary
curvature of Galaxy's field will distort the measured values of parallaxes.
If the observer can restore the parameters of a body distorting the
observational data, then one may perform a reduction procedure and restore
the true position of the light source and its parallax. This procedure,
successfully applied for reduction of observations under the action of bodies
of the solar system, is invalid for taking account of microlensing in the
Galaxy. In most cases  masses and distances to the stars inducing the
nonstationary curvature are unknown. This uncertainty results in the upper
bound of accuracy of the parallaxes being measured.

In particular, the parallax distortions may be so large that the observer
will register a negative value of the parallax instead of a positive one. The
negative value of the parallax of a remote source is usually idenified with
the presence of measurement errors. In fact, the shift of a remote light
source repeatedly measured half a year after the first measurement is a sum
of the parallax shift, proper motion and random error of measurement. If one
considers the proper motion to be equal to zero and the random errors
to have an arbitrary sign (in particular, negative), then the measured
parallax shift will be negative. In this case the total parallax shift is
negative.

In a relativistic case the positive parallax is no longer a necessary
condition of the real observation. Moreover, at a certain level of accuracy
the most part of the observed parallaxes from remote sources will be negative!

The largest distortions of parallaxes will be observed under the action on
light rays from rather close gravitational lenses. When the angular distance
from the lens to the source is comparable with the parallax value, there may
occur a "transfer" of the image from one side of lens to the other side. This
effect will result in a considerable distorsions of the source position and
parallax.

Of course, the weak microlensing effect being observed from the barycentre of
the solar system is not added to the source parallax motion. An attempt to
reduce the position of the source, being observed from the Earth, to the
barycentre following the standard equations will lead to an error since in
this case the parallax shift value itself will depend on the microlensing
effect.

Make a general comment on the whole article.

In the problem related to distortion of measured parallaxes there naturally
appear some quantities being different-order infinitesimals. The unit
vectors indicating the direction to the light source are zeroth-order
quantities.

We shall consider the quantities containing the light source parallax as a
factor to be first-order infinitesimals. As the first-order infinitesimals
we shall also regard the proper shift of the light source or lens for all
time of observation. These values are  products of the proper velocity of
motion of the object by all time of observation. We shall consider the
angular distance between the lens and the light source to be a small
parameter as well.

Besides these small parameters, in the problem there arises one more
small parameter which is not related to the geometry of the problem under
consideration but is related to general relativistic effects. This is
Einstein's-cone-size-to-source-lens-angular-distance ratio squared.

As seen from the foregoing, the small parameters take different values. Thus,
e.g., the lens parallax may be of 10 mas, whereas the parallax of an
extragalactic source may be of 10 nas, which $10^6$ times less.
Nevertheless the terms containing the lens parallax squared are less
than the terms proportional to the first power of the source parallax.
Hence in the equations we shall retain linear terms with respect to
small parameters. Only two exceptions are made. We retain all square
terms, one of the cofactors of which is
Einstein's-cone-to-angular-impact-distance ratio. This small factor
should be retained since the weak microlensing effect is proportional to it,
i.e. the microlensing effect itself (its value) depends on the size of
Einstein's cone. Besides, while calculating
Einstein's-cone-squared-to-angular-distance-squared ratio, we also retain
second-order infinitesimals in the ratio denominator since zeroth- and
first-order quantities are absent there.

The structure of the article is as follows. In Sec. 1 we consider the
equation of a gravitational lens in a vector form suitable for derivating
the observed parallax and solving this equation. Sec. 2 considers the
position measurements performed from two different locations. Then we
consider rigid-baseline parallax measurements. In Sec. 5 we consider a notion
of the parallax ellipse arising in a visible motion of the source on the
celestial sphere for a year as well as discuss characteristic sizes of
the parallax ellipse and a trajectory of the visible displacement of the
source on the sky in the presence of the weak lensing effect.

Finally, in conclusion we discuss the level of accuracy at which distortions
in parallax measurements become significant.

\section*{Equation of a Gravitational Lens and Its Solution}

To derive formulae describing a change in the value of the measured
parallax of a light source, at the presence of a gravitational lens, we
consider the equation of a gravitational lens. We shall consider the
effect only in the case of a weak microlensing \cite{saz96},
\cite{saz98} since in this case we may neglect the second weak image of
a source and consider the first of the images to be unique that we
observe. We shall use the standard model of microlensing based on a
simple model of a point lens with a spherically symmetric gravitational
field. This is the most interesting case since the Galaxy's
nonstationary field, formed by separate stars, is a set of
gravitational fields, with  each of which being spherically symmetric.
While discussing the equation of a gravitational lens, we shall follow
articles \cite{ref64}, \cite{hog} as well as review
\cite{zak98} and books \cite{zak97}, \cite{sch}.

\begin{figure}
\centerline{{\epsfbox{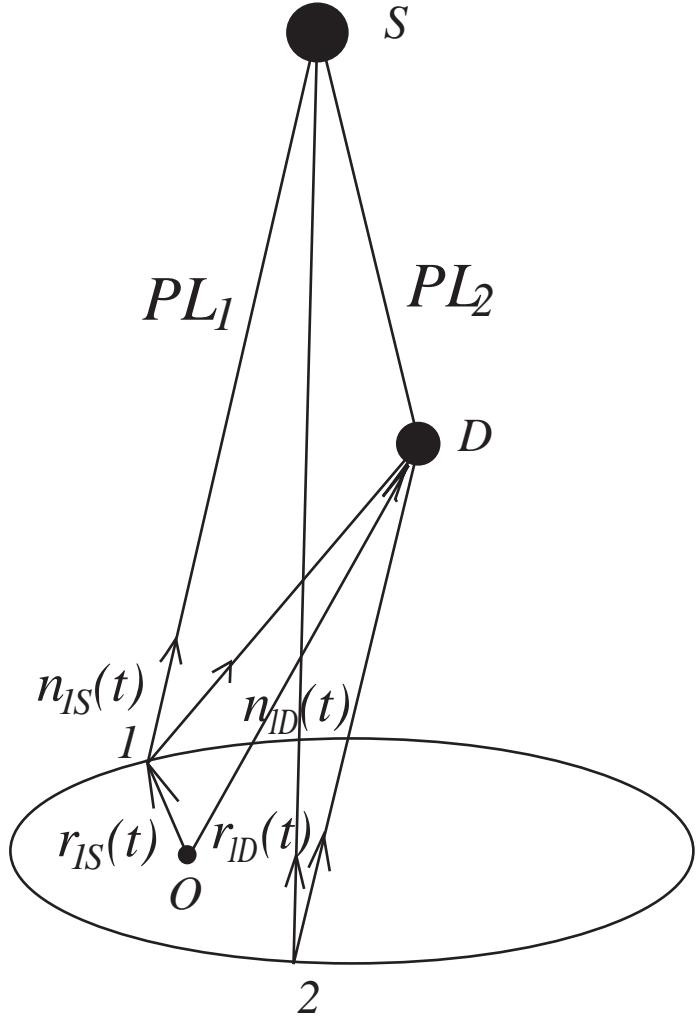}}}
\label{fig1}
\caption{
The figure depicts a position of the light source, the gravitational lens as
well as positions of the observers. The plane $PL_J$ concerns the $J$
observer.
}
\end{figure}

Consider the situation in which there is a light source $S$, a lens $D$
with the mass $M$ and two observers 1 and 2 (designated by $J$). As the
lens $D$ a usual star or a dark body may appear. We shall consider that
the velocity of a proper motion of the bodies $S$, $D$ as well as of
each observer $J$, is much less than the velocity of light. The three
bodies:  the light source $S$, the lens $D$ and one of the observers
$J$ form a plane which may be designated as $PL_J$. Here the index $J$
shows that the plane is formed taking account of the $J$ observer
\ref{fig1}.  The bodies $S$, $D$ and the other observer form a second
plane. It should be noted that the trajectories of photons curved under
the action of the gravitational lens are approximated by broken lines,
as well as in review \cite{zak98}, which makes it possible  for us to
speak about the planes where vectors are located. Besides, the choice
of two planes is also due to it is necessary to take into account
correctly the quantity defined as a difference of angles between the
direction to the lens and the visible position of the source, which
will be discussed below in detail.

The plane $PL_J$ is of importance since the principal vectors of the
problem as well as the light ray trajectory lie in this plane
\cite{zak98}. It is natural that the trajectory of the light incoming
to the other observer belongs to the other plane. Introduce vector
designations in our problem.  Each vector will be regarded at some
instant $t$ considering the position of each of the bodies to be a
function of time. The vector from the origin of coordinate to the light
source $S$ will be denoted as $\vec r_S(t)$. The vector drawn from the
origin of coordinate to the lens will be denoted as $\vec r_D(t)$.
Similarly we draw vectors $\vec r_J(t)$ to each of the observers. The
vector connecting the observer $J$ and the light source is
\begin{equation}
\label{vector1}
\vec r_{JS}(t)= \vec r_S(t)- \vec r_J (t),
\end{equation}
and the
vector connecting the same observer and the lens is
\begin{equation}
\label{vector2}
\vec r_{JD}(t)= \vec r_D (t)- \vec r_J (t)
\end{equation}
The three points $S$, $D$ and the observer $J$ lie in one plane $SDJ$.
The vectors $\vec r_{JS}$ and $\vec r_{JD}$ lie in the same plane. To
write a solution to the equation of a gravitational lens in a vector
form and describe the weak microlensing effect for two spaced
observers, we introduce unit vectors whose directions coincide with
those of the vectors $\vec r_{JS}$ and $\vec r_{JD}$:
\begin{equation}
\vec n_{JS}=\frac{\vec r_{JS}}{r_{JS}}
\end{equation}
and
\begin{equation}
\vec n_{JD}=\frac{\vec r_{JD}}{r_{JD}} .
\end{equation}
Here and below the quantities without vector arrows above designate absolute
values of the vectors $r_{JS}\equiv |\vec r_{JS}|$.

The difference of these two vectors
\begin{equation}
\label{angulardistance}
\Delta \vec n_J = \vec n_{JS} - \vec n_{JD}
\end{equation}
is a vector which belongs to the plane $PL_J$, and is approximately
equal in magnitude to the difference of angles between the directions
to $C$ and $D$.

On separating the plane $S$, $D$ and $J$, the equation of a
gravitational lens reduces to a trivial quadratic equation for angles
\cite{ref64}, \cite{sch}, \cite{zak98}:
\begin{equation}
\theta^2_i -\theta \cdot \theta_i -\theta^2_e=0
\end{equation}
Here $\theta \equiv |\Delta \vec n_J| $ is a difference of the angles between the direction to the lens and
the direction to the true position of the light source, $\theta_i$ is a difference of
the angles to the lens and a visible position of the source (image), $\theta_e$ is
Einstein's cone size:
\begin{equation}
\theta^2_e=\frac{4GM}{c^2}\frac{D_{DS}}{(D_{DS} +D_{DJ})D_{DJ}}
\end{equation}
In the latter formula $M$ is the lens mass, $D_{DS}$ is the distance from the
lens to the light source, $D_{DJ}$ is the distance from the lens to the
observer $J$. It should also be noted that Einstein's cone size differs
for the two observers since $\theta_e$ depends on the distance between the lens
and the observer. However this difference is small and shall be
neglected.

In paper \cite{kop} a nonstationary situation has been analyzed, when an
effect of the motion of the lens on the motion of light rays is taken into
account. From the results of this paper it is seen that in our situation an
effect of nonstationarity of the gravitation field of the lens may be
neglected.

Solving the lens equation, we find two roots $\theta_i$ which correspond to two
image positions of the source. Since we are only interested in the weak
microlensing effect, we shall only be concerned in the position of the main
image. The solution to the equation of a gravitational lens for the position
of the main image is
\begin{equation}
\theta_i={1 \over 2}\theta +{1 \over 2}\theta \sqrt{1+
\frac{4\theta_e^2}{\theta^2}} \simeq \theta + \dfrac{\theta^2_e}{\theta}.
\end{equation}                                                                            .
The difference between the true and visible positions of the source is
\begin{equation}
\delta \theta \simeq \frac{\theta_e^2}{\theta}.
\end{equation}

The image position on the picture plane can be drawn as follows
(see Fig. 2). Through the lens position and the true position of
the source a straight line is drawn. Then one determines a vector whose
origin coincides with the lens $D$ and whose end coincides with the
source position $S$ on the picture plane. The length of this vector is
equal to $\theta$. To find the image position, this vector is
continued in the same direction up to the length $\theta_i$. The
resulting vector is a two-dimensional one determining the image
position.  Necessity of such an image drawing will be clarified below.

\begin{figure}
\centerline{{\epsfbox{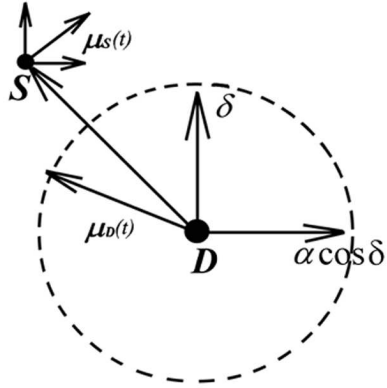}}}
\label{fig2}
\caption{
The figure depicts source and lens positions on the picture plane.
}
\end{figure}

Now one may identify the vector $\vec \Delta n_J$ introduced above with
that connecting the positions $D$ and $S$ on the picture plane. We
denote the vector drawn from the point $D$ to the image $I$ as $\vec
\Delta n_{JI}$, $\theta \equiv |\Delta \vec n_J|$. Now the equation
describing connection between the position and image vectors takes the
form
\begin{equation}
\Delta \vec n_{JI}={1 \over 2}\Delta \vec n_{J} + {1 \over 2}
\Delta \vec n_{J} \cdot \sqrt{1+ \frac{4\theta^2_e}{\Delta n_{J}^2}}
\approx
\Delta \vec n_J + \Delta \vec n_J \dfrac{\theta_e^2}{\Delta n_J^2}
\end{equation}

The visible position of the source relative to the true direction $\vec
n_{JS}$ is expressed as
\begin{equation}
\vec n_{JI}=\vec n_{JS} + \Delta \vec n_J  \frac{\theta^2_e}{\Delta n_{J}^2}
\end{equation}
                                                                            .
The light source $S$, the deflecting body $D$ and the observer $O$ possess a
peculiar motion. We shall consider the source and lens motion to be
rectilinear and uniform. We shall also consider the origin of our
system of coordinates to be related to the barycentre of the solar
system, so that the observer's velocity relative to the selected system
of coordinates is only the velocity of motion about the barycentre.

Let the vector of the three-dimensional velocity relative to the selected
system of coordinates be $\vec v_S$, and the lens velocity be $\vec v_D$.
We shall consider both of these quantities to be constant.

One may divide the three-dimensional velocity into longitudinal and
transversal components. The longitudinal component of the velocity changes
basic physical parameters of the picture under consideration, e.g. such as
Einstein's cone size. However this effect is small and shall be neglected
below. The transverse components of all three motions are added to result in
a mutual motion of the source $S$ and the body $D$ in the
observer's picture plane.

Since the velocities of bodies are constant, and the observer's velocity
relative to the barycentre is given, now we can calculate a law of variation
in the vectors $\vec n_{JS}$, $\vec n_{JD}$ as well as other vectors required for solving our main
problem of calculating the image motion, with the observer's parallax motion
and the uniform motion of the source and lens taken into account.

Fix the source and lens positions at the instant $t=0$. Then the vector
connecting the observer $J$ and source satisfies the equation
\begin{equation}
\vec r_{JS}(t)=\vec r_{S0} +\vec v_S t -\vec r_J(t)
\end{equation}

Below the values at the instant $t=0$ will be everywhere designated by
the index 0.

\noindent
The unit vector in the direction from the observer $J$ to the source is
described by the equation
\begin{equation}
\vec n_{JS}(t)=\vec n_{S0} +(\dfrac{\vec v_S}{r_{S0}}
-\vec n_{S0}\dfrac{(\vec n_{S0} \vec v_S)}{r_{S0}})t
-(\dfrac{\vec r_J(t)}{r_{S0}} -\vec n_{S0} \dfrac{(\vec n_{S0} \vec r_J)}{r_{S0}})
\end{equation}

Consider two vectors: the first one
\begin{equation}
\frac{\vec v_S}{r_{S0}}- \vec n_{S0}\frac{(\vec n_{S0} \vec v_S)}{r_{S0}},
\end{equation}
and the second one
\begin{equation}
(\dfrac{\vec r_J(t)}{r_{S0}} -\vec n_{S0} \dfrac{(\vec n_{S0} \vec r_J)}{r_{S0}}).
\end{equation}

These are three-dimensional vectors perpendicular to the vector $\vec n_{S0}$. Hence
from the viewpoint of the first observer, at the instant $t=0$, they are
two-dimensional vectors lying in the picture plane. Besides, the first vector
has dimensions s$^{-1}$ and coincides with the proper angular velocity
of displacement of the source on the sky. Hence we denote the first of
these vectors by the two-dimensional vector $\vec \mu_S$
\begin{equation}
\vec \mu_S =\frac{\vec v_S}{r_{S0}}- \vec n_{S0}\frac{(\vec n_{S0} \vec
v_S)}{r_{S0}},
\end{equation}
and the second one will be denoted as
\begin{equation}
\vec \beta_{JS}= \dfrac{\vec r_J(t)}{r_{S0}} -\vec n_{S0} \dfrac{(\vec n_{S0} \vec r_J)}{r_{S0}}
\end{equation}

It should be also noted that the vector $\vec \mu_S$, being small in magnitude, is
equal to $|\mu| \sim  10^{-11}$ s$^{-1}$ for the fastest and closest
objects, and is small as $|\mu| \sim 10^{-20}$ s$^{-1}$ for object at
cosmological distances.

Similarly one can calculate the unit vector from the observer to the lens.
\begin{equation}
\vec n_{JD}(t)=\vec n_{D0} + \vec \mu_D t
-\vec \beta_{JD}(t)
\end{equation}

It is in this case that we should introduce new designations, with the
angular velocity of the lens being determined by the vector $\vec n_{D0}$
\begin{equation}
\vec \mu_D =\frac{\vec v_D}{r_{D0}}- \vec n_{D0}\frac{(\vec n_{D0} \vec
v_D)}{r_{D0}},
\end{equation}
and the second vector satisfies the equation
\begin{equation}
\vec \beta_{JD}= \dfrac{\vec r_J(t)}{r_{D0}} -\vec n_{D0} \dfrac{(\vec n_{D0} \vec r_J)}{r_{D0}}.
\end{equation}
Now the index $D$ denotes the corresponding vectors for the lens.

The vector $\vec \mu_D $ is the angular velocity of the lens, and the
vector $\vec \beta_{JD}$ has the same meaning as the vector $\vec
\beta_{JS}$ drawn from the observer to the source.

Both vectors $\vec \mu_D$ and $\vec \beta_{JD}$ are perpendicular to
the vector $\vec n_{D0}$. They are first order infinitesimals.  Recall
that the vector $\vec n_{S0}$ differs from the vector by a first-order
infinitesimal $\Delta \vec n_0$.  Hence the vector $\vec \mu_D$ may be
also considered perpendicular to the vector $\vec n_{S0}$ to a
second-order infinitesimal, and the equation can be written by
substituting the vector $\vec n_{D0}$ by the vector $\vec n_{S0}$
referring it to one of the basic vectors of our problem $\vec n_{S0}$.

The vector $\vec \beta_{JD}$ differs from the vector $\vec \beta_{JS}$
by a first-order infinitesimal
\begin{equation}
\delta \vec \beta_J= \vec \beta_{JD} -\vec \beta_{JS} =
(\frac {1}{r_{D0}} - \frac{1}{r_{S0}} ) \lbrack \vec r_J -
\vec n_{S0}(\vec n_{S0} \vec r_J)\rbrack,
\end{equation}
although the scalar product of $\vec \beta_{JD}$ and $\vec n_{S0}$ is
equal to zero to a second-order infinitesimal. Hence we shall consider
the vectors $\vec \beta_{JD}$ and $\vec n_{S0}$ to be mutually perpendicular.

Finally, we define a vector of the angular difference between true directions
to the light source and to the lens at the instant $t=0$
\begin{equation}
\Delta \vec n_0=\vec n_{S0} -\vec n_{D0},
\end{equation}
a difference of angular velocities of the source and the lens respectively
\begin{equation}
\vec \mu =\vec \mu_S - \vec \mu_D.
\end{equation}
Then we obtain a dependence of the vector $\Delta \vec n_J(t)$ in the form
\begin{equation}
\Delta \vec n_J(t)= \Delta \vec n_0 +\vec \mu t +\delta \vec \beta_J(t).
\end{equation}

Finally, we write down the equation for an image position
\begin{equation}
\vec n_{JI} = \vec n_{SO} + \vec \mu_S t - \vec \beta_{JS}(t) +
(\Delta \vec n_0 + \mu + \delta \beta_{JS}),
\frac {{\theta_e}^2}{\Delta n_J^2(t)}
\end{equation}
where
\begin{equation}
\Delta n_{J}^2(t) = \Delta n_0^2 + \mu^2 t^2 + \delta \beta_J^2 +
2(\Delta \vec n_0 \vec \mu ) t + 2( \Delta \vec n_0
\delta \vec \beta_J (t) ) t.
\end{equation}

In this equation second-order infinitesimals are retained since, as mentioned
in the introduction, zero- and first-order terms are absent in this sum.

\section*{Measurements from Two Positions}

Celestial source position measurements performed from two spaced points
allow a valuable additional information on this source to be obtained. Of
prime importance is the measurement of a trigonometric parallax of the
source, which allows its distance to be measured.

The parallax measurements may be carried out by several ways \cite{gre},
\cite{mar}, \cite{eic}. We consider the measurements of interferometric type,
being conducted by observers located at different ends of a rigid baseline,
as well as the parallax motion observation conducted by the observer located
on the moving Earth. The theory of such observations in Euclidean space may
be found in manuals on astrometry. There arises a special feature in the
presence of the weak microlensing effect. The plane $PL_J$ moves in
space.  This occurs mainly due to the observer's moving about the
barycentre of the solar system.

In such measurements, especially in the second case, of importance is a
mutual orientation of the baseline and the picture plane. On this the
arc $\theta_i$ depends, which makes it necessary to determine the
image position via a two-dimensional vector. This permits a change in
the arc $\theta_i$ to be taken into account correctly, while passing from the
first source plane to the second one.

\begin{figure}
\centerline{{\epsfbox{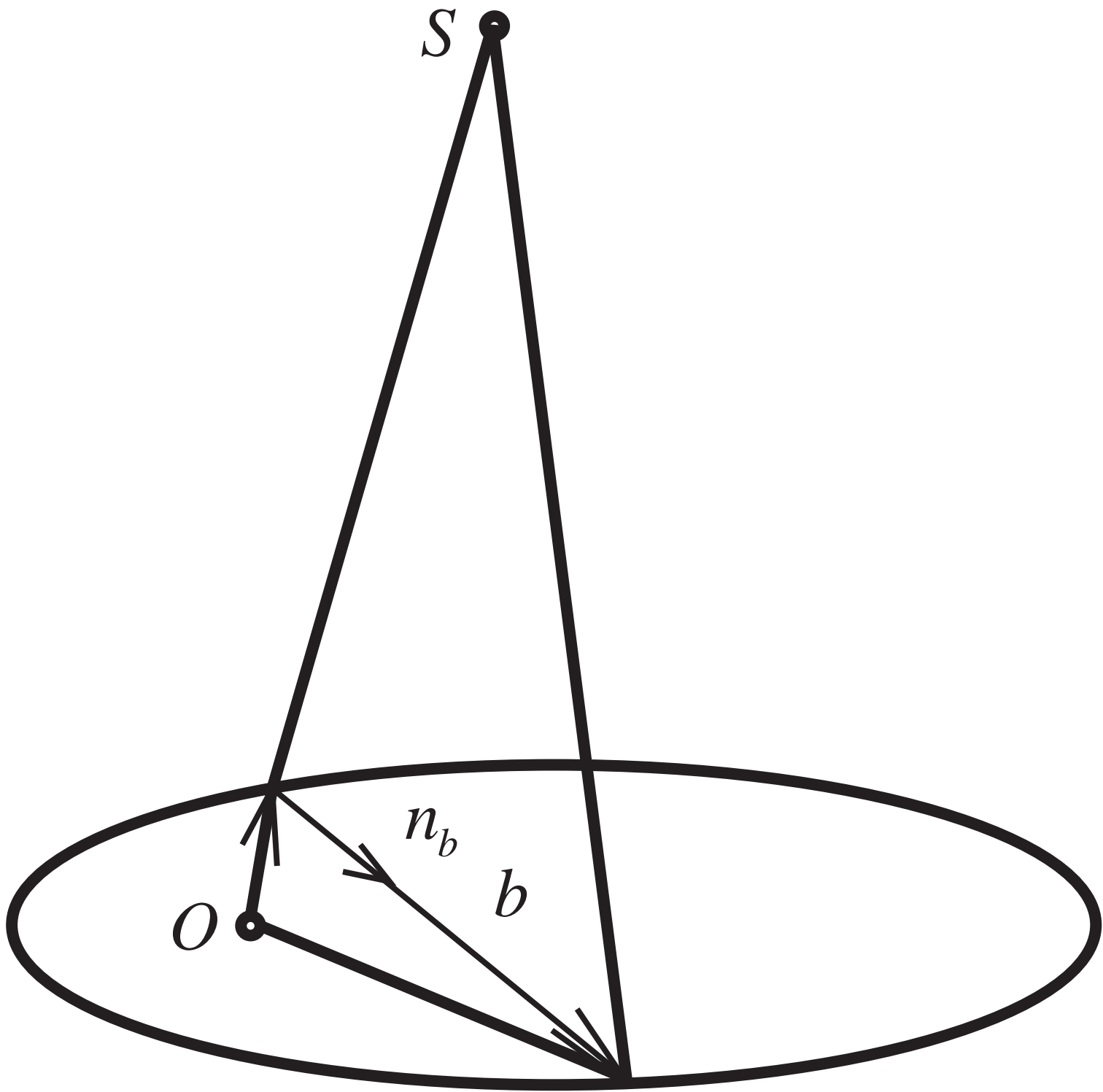}}}
\label{fig3}
\caption{
The baseline vector is defined as a difference of the vectors drawn to the
first and second observers respectively.
}
\end{figure}

Introduce a vector directed from one point of observation to the other, which
we shall call the baseline vector $\vec B$. Then the difference of the vector
drawn from the origin of coordinates to the first observer and that to the
second one is (see Fig. 3 \ref{fig3})
\begin{equation}
\vec B =\vec r_2(t_2)- \vec r_1(t_1).
\end{equation}
As seen from this definition, the baseline vector is generally a function of
time. Define also a vector coinciding the baseline direction
\begin{equation}
\vec n_b = \frac {\vec B}{B}
\end{equation}

Now discuss the basic equations describing observations from two spaced
points. At first instant $t$ the vectors, one of which connects the
first observer and the source, and  the second one connects the first
observer and the lens, are $r_{1S}(t_1)$ and $r_{1D}(t_1)$. The second observer is located at
another point, and its position differs from the position of the first
observer by a vector $\vec B$.  Photons, generally speaking, arrive to the
observer at the instant differing from $t$ by the value $\sim \dfrac
{B}{c}$. Therewith we should calculate source and lens positions at
another instant.  However this difference is small and shall be
neglected.

The vector connecting the source and the second observer can be written as
\begin{equation}
\vec r_{2S} = \vec r_{1S}-\vec B ,
\end{equation}
and the true direction to the source, for the second observer, is
\begin{equation}
\vec n_{2S}=\vec n_{1S} - \frac{B}{r_{1S}}  (\vec n_b - \vec n_{1S}(\vec n_{1S}\vec n_b))
\end{equation}
We shall consider that the distances from the observers to the origin of
coordinate system are much less than the distance from the source to the
observers. Then we substitute $r_{1S}$ by $r_{SO}$ and, neglecting
small quantity squares, we obtain
\begin{equation}
\label{vector}
\vec n_{2S}(t) = \vec n_{1S}(t) - \frac {B}{r_{SO}} \vec b,
\end{equation}
where
\begin {equation}
\vec b = \vec n_b  - \vec n_{SO}(\vec n_{SO} \vec n_b)
\end{equation}
A similar equation is valid for the vectors connecting the observer and the
lens
\begin{equation}
\vec r_{2D}(t)= \vec r_{1D}(t)- \vec B ,
\end{equation}                                                                          ,
and for the unit vectors in the lens direction
\begin{equation}
\vec n_{2D}= \vec n_{1D}- {\frac{B}{r_{1D}}}[ \vec n_b- \vec n_{1D}(\vec n_{1D} \vec n_b)]
\end{equation}
Transform the equation for $\vec n_{2D}$ similarly to the equation
(\ref{vector}) dropping second-order infinitesimals and singling out
$r_{DO}$ and $\vec n_{SO}$
\begin{equation}
\vec n_{2D} = \vec n_{1D}- \frac {B}{r_{DO}}\cdot \vec b
\end{equation}
Now we find relation between the vectors $\Delta \vec n_2$ and  $\Delta
\vec n_1$ connecting the light source and the lens in the picture
planes of the second and first observers.  These vectors enters into
the denominator of the factor determining the weak microlensing effect
\begin{equation}
\label{vct}
\Delta \vec n_2 = \Delta \vec n_1 - B \left( \frac {1}{r_{SO}} - \frac {1}{r_{DO}}
\right ) \vec b
\end{equation}
In  equation (\ref{vct}) we have retained only first-order infinitesimals
both in the left and right side of the equation. Introduce the designations
$p_S = \dfrac {B}{r_{SO}}$ and $p_D = \dfrac {B}{r_{DO}}$ for the light source and lens parallaxes respectively as well as
for $p = p_S - p_D$. We present the final equation for the vector directed to the source
image from the second observer's position
\begin{equation}
\label{equation2}
\vec n_{2I}=\vec n_{2S}+\Delta \vec n_2 \frac{{\theta_e}^2}{\Delta
{n_2}^2}
\end{equation}
and from the first observer's position
\begin{equation}
\label{equation1}
\vec n_{1I}=\vec n_{1S}+\Delta \vec n_1 \frac{{\theta_e}^2}{\Delta
{n_1}^2}.
\end{equation}
The angular distances are connected by the equality
\begin{equation}
\label{equation3}
\Delta \vec n_2=\Delta \vec n_1-p \vec b,
\end{equation}
and the vectors showing the true position of the source depend on time as
\begin{equation}
\label{equation4}
\vec n_{1S}=\vec n_{SO}+\vec \mu_{S}t-\vec \beta_{1S}(t),
\end{equation}
\begin{equation}
\label{equation5}
\vec n_{2S}=\vec n_{SO}+\vec \mu_{S}t-\vec \beta_{2S}(t).
\end{equation}
and differ only by a parallax vector $p_S \vec b $.

\section*{Rigid-Baselined Observations}

We shall call measurements rigid-baselined providing that B=const. So is
measured a distance to some object in geodesy. At a constant baseline
conducted are VLBI observations for which characteristics of the arriving
rays are of importance as well. While observing from the Earth's surface, the
VLBI baseline varies due to tides, tectonic shears and other processes. The
value of these variations of the order of 50 cm per day. In our problem these
variations of the baseline may be neglected, and the baseline may be
considered constant. Besides, this case is valuable from a methodical
viewpoint since it allows one to clarify a physical sense of the quantities
being measured as well as a mechanism of visible change in the parallax. As
well as everywhere, we shall consider the rays to arrive at both ends of the
baseline simultaneously, i.e. the variation in the basic vectors of our
problem for the time interval $\dfrac{B}{c} $ is negligible. Besides, in this section we
shall consider $p_S<<p_D$.

While measuring the parallaxes, one tries to orient the baseline so that one
of its ends, say, the first were directed perpendicular to the source image
\begin{equation}
\label{cond1}
(\vec n_{1I} \vec n_b)=0
\end{equation}
In this case the instant t=0 is chosen as a beginning of observation,
and the origin of coordinates is assumed to coincide with the first
observer. Then $\vec n_{1S}=\vec n_{SO} $. In Euclidean geometry the direction to the image is
a true direction to the source, thus the perpendicularity condition is
reformulated as
\begin{equation}
\label{cond2}
(\vec n_{SO} \vec n_b)=0,
\end{equation}
and the vector $\vec \beta_{2S} $ becomes a vector $ p_{S} \vec
\beta$.

In the absence of the weak microlensing effect light rays move along straight
lines. Now the scalar product takes the form (\ref{cond1})
\begin{equation}
(\vec n_{2S} \vec n_b)=-p_S
\end{equation}

Define the visible value of the parallax as a scalar product of the
vectors $\vec n_{1S}$ (directed to the image from the second
observer's position) and $\vec n_b$ taken with opposite sign. This definition
is equivalent to the parallax definition in Euclidean astrometry.
\begin{equation}
\label{cond3}
p_a=-(\vec n_{2I} \vec n_b)
\end{equation}

The condition (\ref{cond1}) together with the equation
(\ref{equation1}) for the weak microlensing effect determines the
equation
\begin{equation}
\label{equation6}
(\vec n_{1S} \vec n_b)=- \frac{(\Delta \vec n_1 \vec n_b)}{\Delta
n_1} \frac{{\theta_e}^2}{\Delta n_1}
\end{equation}

\noindent
Introduce an angle between the baseline vector $\vec n_b$ and the
angular-impact-parameter vector $\Delta \vec n_1$ in the form
\begin{equation}
\label{cond4}
\frac{(\Delta \vec n_1 \vec n_b)}{\Delta n_1}= \cos \psi
\end{equation}
Now we calculate the parallax value measured according the condition
(\ref{cond3}) and equations (\ref{equation2}), (\ref{equation5}),
(\ref{equation3})
\begin{equation}
\label{equation7}
p_a=p_s+p(\vec b \vec n_b)\frac{{\theta_e}^2}{\Delta
{n_2}^2}+(\Delta \vec n_1 \vec n_b) \left(
\frac{{\theta_e}^2}{\Delta {n_1}^2}-\frac{{\theta_e}^2}{\Delta
{n_2}^2} \right)
\end{equation}

The projection onto the baseline vector differs from unity by second-order
infinitesimals to be neglected. Hence we shall assume that $(\vec b
\vec n_b)=1$. The source and lens parallax difference $p=p_S-p_D$ is
multiplied by a small factor $\dfrac{{\theta_e}^2}{\Delta {n_2}^2}$.
For close sources, whose parallax is comparable with the lens parallax,
the weak microlensing effect will not change the parallax too much.
However in case $p_S<<p_D$ and $p_S \sim p_D
\dfrac{{\theta_e}^2}{\Delta {n_2}^2}$ a change in the visible parallax
may be significant.  We shall also consider that the angular distance
between the source $S$ and the lens $D$ is much more than the
parallaxes.  In this case the third term is comparable with the second
one in magnitude. Of course, $(\Delta \vec n_1 \vec
n_b) \dfrac{{\theta_e}^2}{\Delta {n_1}^2}$ and $(\Delta \vec n_1 \vec
n_b) \dfrac{{\theta_e}^2}{\Delta {n_2}^2}$ far exceeds both the
first and second terms in equation (\ref{equation7}).  However their
difference is already comparable with the second term of equation
(\ref{equation7}).  We resort to the definition (\ref{cond4}) and
obtain the visible value of the parallax
\begin{equation}
\label{ equation8}
p_a=p_s+(p_D-p_S) \frac{{\theta_e}^2}{\Delta {n_2}^2} \cos 2\psi
\end{equation}

In equation (\ref{equation8}) we neglect terms containing the factor $\sim p^2$.

\begin{figure}
\epsfxsize=\hsize
\centerline{{\epsfbox{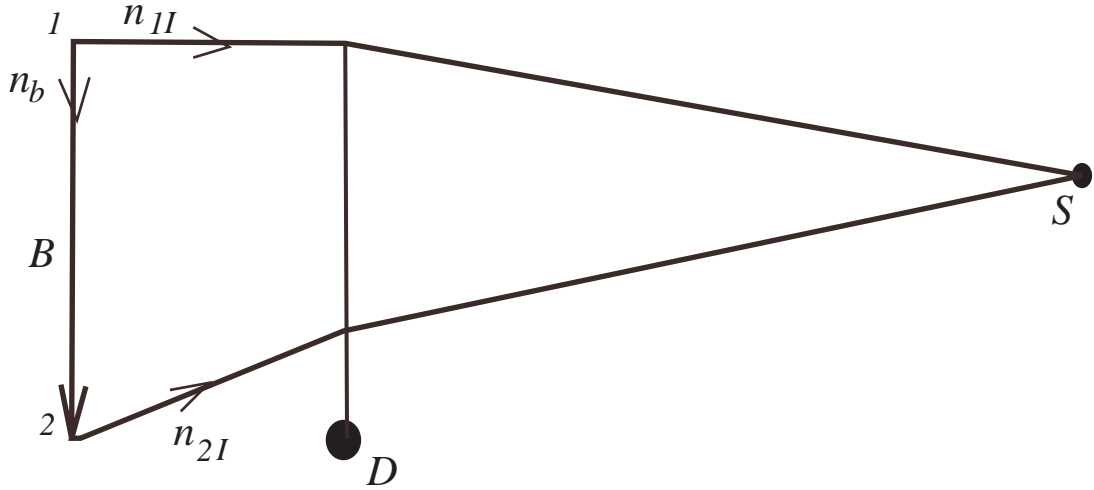}}}
\label{fig4}
\caption{
The lens, baseline and light source are located so that the lens lies off the
triangle formed by two observers and the source.
}
\end{figure}

Consider more fully two situations related to different arrangements of the
baseline vector relative the vector of the angular distance between the
source $S$ and the lens $D$. The first situation arises when the
baseline vector, the lens $D$ and the source $S$ lie in the same plane,
with the lens $D$ lying off the triangle formed by the baseline and two
rays drawn from the source $S$ to different ends of the baseline (see
Fig. \ref{fig4}). In this case $\psi=0$, and the visible parallax is a sum
of the true parallax and an additional term
\begin{equation}
p_a=p_S+p_D \frac{{\theta_e}^2}{\Delta {n_2}^2}
\end{equation}

The visible parallax will be more than the true one. The vectors directed
from the ends of the baseline to the source image shall be "shifted" under
the action of the gravitational field of the lens. Such a disposition, the
baseline-lens-source, will imitate an "approach" of the source to the
observer.

An increase in the visible parallax is valid for the situation when the lens
$D$ does not belong to the plane formed by the baseline $\vec B$ and the
source.  However an increase in $p_a$ is valid only for the angles $\psi$ less
than $\dfrac{\pi}{4}$. It should be emphasized the the value of the
vector $\Delta n_1$ is more than the lens parallax $p_D$, i.e. $\Delta
n_2$ has the same direction as $\Delta n_1$.

The second situation arises when the lens does not belong to the lens formed
by the baseline $(\vec B)$ and the source. In case $\cos 2 \psi=-1$,
i.e. the baseline vector is perpendicular to the plane $(\vec B S)$,
the visible parallax diminishes
\begin{equation}
p_a=p_S-p_D \frac{{\theta_e}^2}{\Delta {n_2}^2}
\end{equation}
Fig. 5 depicts locations of the baseline $\vec B$, the lens $D$ and the
source $S$ corresponding to this situation. Of cource, the lens $D$ is
assumed to be located under the plane $(\vec B S)$. Depending on the
value of the true parallax $p_S$ and a gravitational additional term, the
visible parallax may be even negative.

\begin{figure}
\epsfxsize=\hsize
\centerline{{\epsfbox{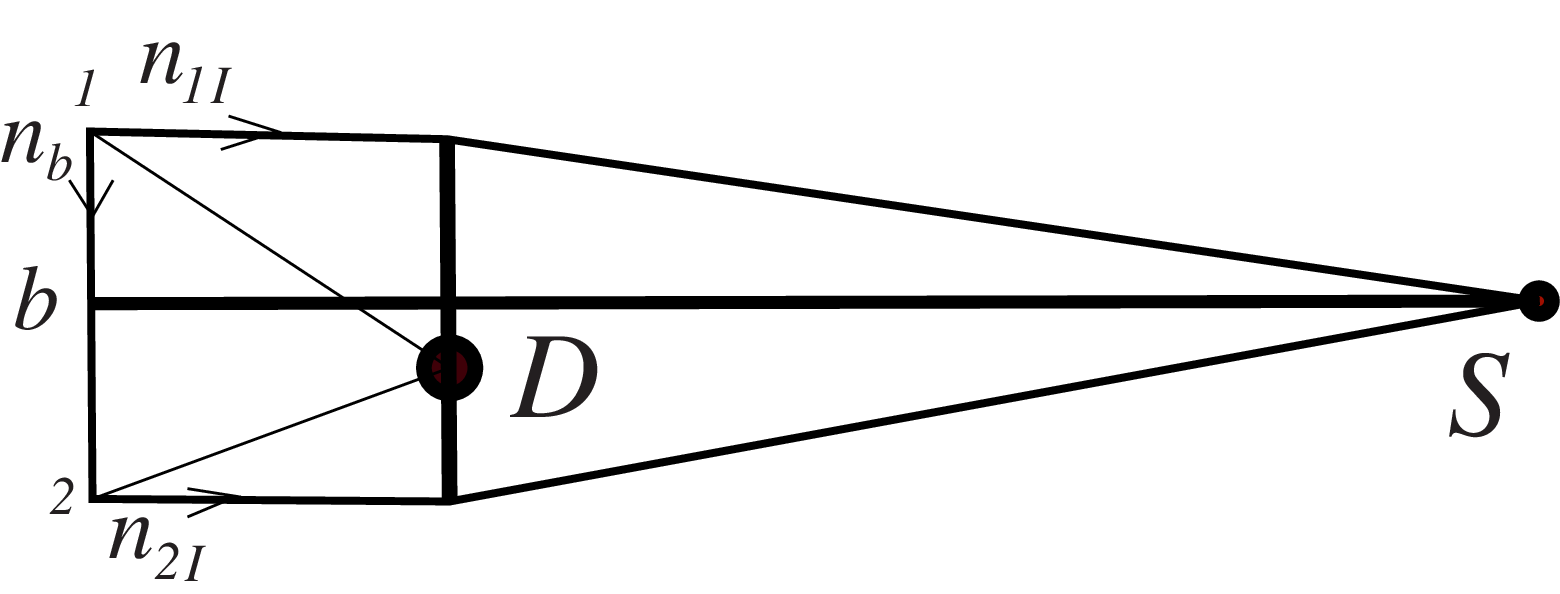}}}
\caption{
The case corresponding to the second variant of locations of the source, lens
and baseline.
}
\label{fig5}
\end{figure}

Of special interest is the situation when the lens $D$ belongs to the
plane $(\vec B S)$, but is located inside the triangle formed by the
baseline and rays of the source $S$, as in Fig. 5, in case $D$ is in
the figure plane.  In this case the parallax is more than $\Delta
n_1$, the terms squared in $p$ become more than the product of
the parallax by the angular impact parameter, and there arises the
second situation in spite of the lens lying in the plane $(\vec B S)$. The
vectors directed from the ends of the baseline to the source image will
be "moved apart" by the gravitational field of the lens $D$. In case
the lens $D$ belongs to the plane $\vec B S$, and the angular impact parameter
$\Delta n_1 \sim p_d$, a small mass as the gravitational lens suffices for the observed
value of the parallax to become zero $p_a=0$ at
\begin{equation}
m_D \sim \frac{c^2}{4G} \frac{B^2}{r_S}
\end{equation}

A direction of the vector $\vec n_{2I}$ should be mentioned. In relativistic
measurements of the paralax of a source the vectors $\vec n_{1I}$,
$\vec n_{2I}$ and $\vec n_b$ lie in the plane formed by three points: observers
1 and 2 as well as the source $S$.

If the curvature of space-time is taken into account, light rays move along
curved trajectories, and the vectors $\vec n_{1I}$, $\vec n_{2I}$ and
$\vec n_b$ do not already belong to one plane. Using the condition
$(\vec n_b \vec n_{1I})$, we form the plane wherein lie the vectors $\vec n_b$
and $\vec n_{1I}$. The equation determining the visible parallax singles out only
one of the three components of the vector $\vec n_{2I}$. The basic
component $\sim$ is directed to the source image and does not contain
additional information.  There exists the third component perpendicular
the plane formed by the vectors $\vec n_b$ and $\vec n_{1I}$. This
component has no analogue in nonrelativistic optics and is equal in
magnitude
\begin{equation}
2(p_D -p_S)\frac{\theta^2_e}{\Delta n_1^2} \frac{(\Delta \vec n_1 \vec
n_b)}{\Delta n_1^2} \left( \Delta \vec n_1 -\vec n_b (\Delta \vec n_1
\vec n_b) \right)
\end{equation}

Generally speaking, this component may be a source of additional information,
e.g., on the measurements conducted being distorted by the weak microlensing
effect.

\section*{Annual Parallax Measurements}

In paper \cite{saz98} a change in the visible position of a light source is
considered in the observer's uniform motion relative the lens and the light
source itself. Although not indicated, in the above paper it was  assumed
that the observations are conducted from the barycentre of the solar system
which moves uniformly and rectilinearly to a first approximation.

While measuring an annual parallax shift, it is natural for the coordinate
system centre to be also chosen coinciding with the barycentre of the solar
system. Then the light source velocity $S$ and the lens velocity $D$
will be differences of the velocities $S$ and $D$ respectively and
those of the barycentre of the solar system. Again we shall consider
that these velocities are constant. Now the baseline is the Earth's
radius vector relative the solar system barycentre. $\vec B$ satisfies a
Keplerian motion.

Clealy the visible motion of the source will already be nonuniform at a
nonzero parallax even in the absence of lensing effect as seen from the
Earth executing the annual motion around the Sun. Define the vector
$\vec n_{2S}$ as a vector directed to the source from the Earth, and
the vector $\vec n_{1S}$ as a vector directed to the source from the
solar system barycentre. These two vectors are connected via the
parallax deviation vector
\begin{equation}
\vec n_{2S}=\vec n_{1S} -p_S \vec b,
\end{equation}
a similar equation can be written to connect unit vectors to the lens both
from the solar system barycentre and the observer on the Earth
\begin{equation}
\vec n_{2D}=\vec n_{1D} -p_D \vec b.
\end{equation}

In the presence of a lens the light source motion being observed from the
barycentre is described by a unit vector of the form
\begin{equation}
\label{lens}
\vec n_{1I}=\vec n_{1S} +\Delta \vec n_1 \frac{\theta^2_e}{\Delta n_1^2}.
\end{equation}

Here $\Delta \vec n_1$ is a difference of unit vectors directed to the
source $\vec n_{1S}$ and to the lens $\vec n_{1D}$. We shall consider
that the motion of the source on the sky is described by a linear
equation of the form
\begin{equation}
\label{motions}
\vec n_{1S}=\vec n_{0S} + \vec \mu_S t,
\end{equation}
and the lens motion -- by the equation as follows
\begin{equation}
\label{motionl}
\vec n_{1D}=\vec n_{0D} + \vec \mu_D t
\end{equation}

The difference vector $\Delta \vec n_1$ is also a linear
time-dependent function
\begin{equation}
\label{dmotions}
\Delta \vec n_1= \Delta \vec n_0 + \vec \mu t
\end{equation}
Here $\vec \mu$ is a difference of angular velocities of the source and the lens, and
$\Delta \vec n_0$ is a difference of initial positions.

Now the source motion observed from the barycentre has the form
\begin{equation}
\label{lenmot}
\vec n_{1I}=\vec n_{0S} + \vec \mu_S t + (\Delta \vec n_0 + \vec \mu t)
\frac{\theta^2_e}{\Delta n_1^2}
\end{equation}

\begin{figure}
\epsfxsize=0.7\textwidth
\centerline{\fbox{
\epsfbox{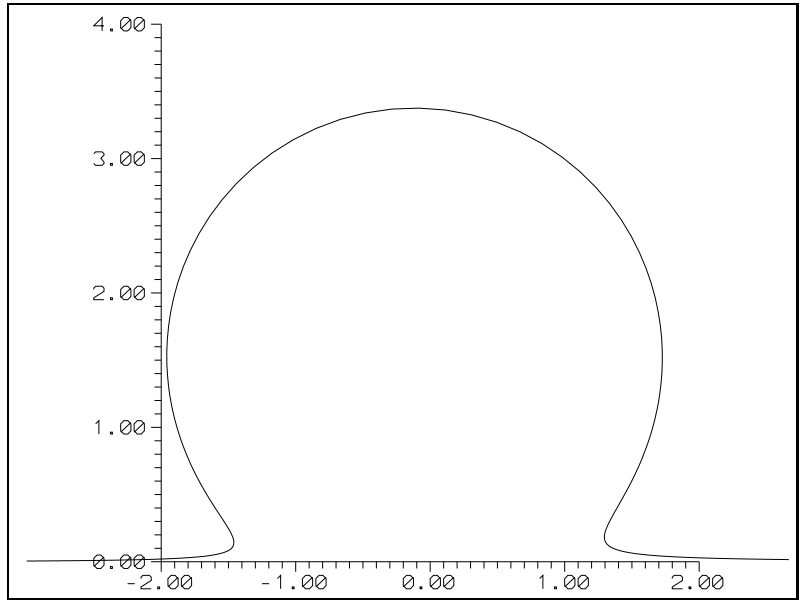}}}
\caption{
The trajectory of a visible motion of the source. In this figure the
direction of the proper motion of the lens is opposite to the source motion
direction.
}
\label{fig6}
\end{figure}

\begin{figure}
\epsfxsize=0.7\textwidth
\centerline{\fbox{
\epsfbox{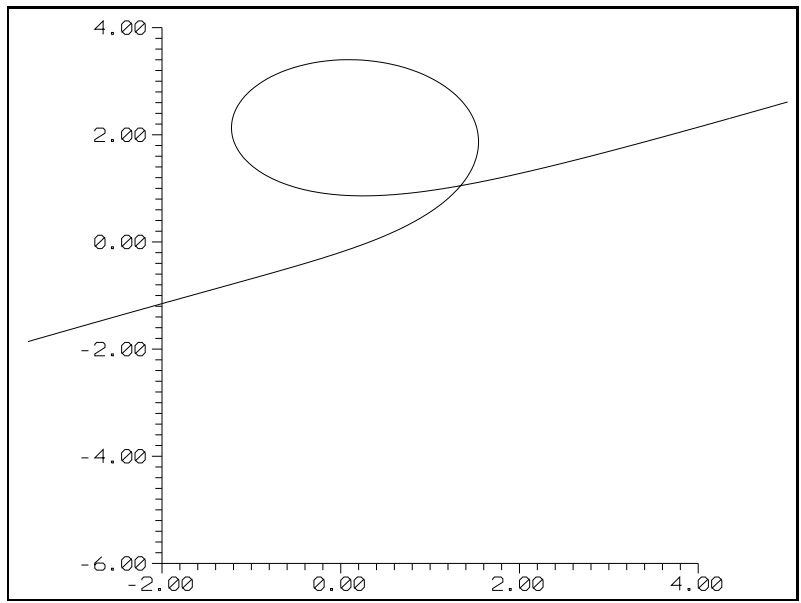}}}
\caption{
The trajectory of a visible motion of the source in the second case of the
proper motion of the source coinciding with the proper motion of the lens.
}
\label{fig7}
\end{figure}

The visible displacement of the source due to the weak microlensing effect
found in \cite{saz98} was calculated at a zero proper motion of the source,
which is valid for most extragalactic objects. In the case of a nonzero
proper motion of the source its motion picture will differ from the case
shown in \cite{saz98}. Now the motion trajectory will have the shape of a
nonclosed curve (see Fig. \ref{fig6}) or an intersecting curve (Fig. \ref{fig7}).

In the absence of a gravitational lens the source position from the Earth's
observer viewpoint is described by an equation of the form
\begin{equation}
\label{earthposition}
\vec n_{2S}=\vec n_{1S} -p_S \vec b.
\end{equation}
When the gravitational lens distorts the source position, the equations of
motion will become more complex
\begin{equation}
\label{lenmotearth}
\vec n_{2I}=\vec n_{0S} + \vec \mu_S t - p_S \vec b+ \Delta \vec n_2(t)
\frac{\theta^2_e}{\Delta n_2^2(t)}
\end{equation}
where $\Delta \vec n_2=\Delta \vec n_1 -p\vec b$

The form of the equation of motion in celestiall coordinates as well as  the
figures illustrated it will be cosidered in the next section.

\section*{Measurement of Coordinates and Source Parallaxe}

In the previous sections we have developed a general theory of observations,
written in a vector form, at two instants from two rigidly bound space
positions as well as discussed parallax measurement from the moving Earth in
the presence of the weak microlensing effect. However in observational
astronomy it is more common to use equations writted in astronomical
coordinates, but not in a vector form. It is of importance since studying the
motion of celestial sources is related at any rate to precalculating the
source positions at a given instant in a certain coordinate system. Some
systems of astronomical coordinates are used, in this case we shall use an
ecliptic system $\lambda$, $\beta$.

In modern astrometric catalogues (HIPPARCOS, TYCHO) the position and
motion of a star is characterized by five parameters chosen in a
certain epoch and referred to a certain place. The choice of the epoch
is of no importance for us. But we shall choose $\lambda$ and $\beta$
as coodinates referred to the solar system barycentre. Two coordinates
of five point to the source. The Cartesian components of the unit
vector in the direction determined by the coordinates $\lambda$ and
$\beta$ are representable as
\begin{equation}
\vec n =(\cos \lambda \cos \beta, \sin \lambda \cos \beta, \sin \beta)
\end{equation}
Two more parameters, presented in the catalogues, are the proper motion
in $\lambda$ and $\beta$, with the velocity with respect to the coordinate
$\lambda$ being a time derivative of the coordinate
\begin{equation}
\mu_{\beta}=\frac{d \beta}{dt}
\end{equation}
and the velocity with respect to the right ascension is defined as
\begin{equation}
\mu_{\lambda}=\frac{d \lambda}{dt} \cos \beta
\end{equation}
one can define the total angular velocity vector as
\begin{equation}
\vec {\mu}= \vec p \mu_{\lambda}+ \vec q \mu_{\beta}
\end{equation}
where $\mu_{\lambda}$ and $\mu_{\beta}$ have been defined above, and
the vectors $\vec p$ and $\vec q$ are
\begin{equation}
\vec p =(-\sin \lambda, \cos \lambda, 0 )
\end{equation}

\begin{equation}
\vec q =(-\cos \lambda \sin \beta, -\sin \lambda \sin \beta, \cos
\beta)
\end{equation}

The vectors $\vec n$, $\vec p$ and $\vec q$ form a triad of mutually
perpendicalar vectors as it is usually defined in astronomy \cite{mar},
\cite{hip}.

Since we shall refer our observations to the solar system barycentre, the
source coodinates will be set in the ecliptic coordinate system, and as the
vector $\vec n_1$ we choose the direction to the source from the solar system
barycentre. Since the first observer is always located at the solar system
barycentre, the vector $\vec B$ is a vector connecting the barycentre and the
observer, everywhere below we shall consider the observer to be located on
the Earth, and the vector $\vec B$ to connect the solar system barycentre with
the Earth centre.

Thus the vector $\vec n_b$ has the form
\begin{equation}
\vec n_b =(\cos \lambda_E, \sin \lambda_E, 0),
\end{equation}
here $\lambda_E$ is the Earth's right ascension in the ecliptic
coordinates from the solar system barycentre. Clearly the Earth's
declination is equal to zero.

Fix the vector triad at the instant $t=0$ and denote its vectors by
index "0":  $\vec n_{S0}$, $\vec p_0$, $\vec q_0$. The ecliptic
coordinates of the place will be also denoted by index "0". We shall
reckon the image position from the true position of the source at the
zero instant. This difference is representable in the form
\begin{equation}
\label{divgeneral}
\vec n_{2I} - \vec n_{S0} = \Delta \lambda (t) \cos \beta_0 \vec p_0  +
\Delta \beta (t) \vec q_0
\end{equation}

\noindent
It is easy to to notice that this equality determines an approximate
expansion of the vector $\vec n_{2I}$ in the triad of perpendicular
vectors $\vec n_{S0}$, $\vec p_0$, $\vec q_0$. An exact expansion of
the vector in this triad differs from the approximate one by terms
containing the factors $\sim \Delta \lambda^2$, $\sim \Delta \beta^2$
to be neglected.

The difference of the unit vectors directed to the source and the lens
(\ref{dmotions}) is representable in the form
\begin{equation}
\label{66}
\Delta \vec n=\Delta \lambda \cos \beta_0 \vec p_0 +\Delta \beta \vec q_0.
\end{equation}
Here $\lambda_0$, $\beta_0$ are coordinates of the light source
position $S$ in the epoch $t=0$, $\Delta \lambda= \lambda_S- \lambda_D$
are differences of the right ascension of the light source and the lens
$D$, and $\Delta \beta= \beta_S-\beta_D$ are differences of the light
source and lens declinations.

Now introduce a new definition $x=\Delta \lambda \cos \beta_0$,
$y=\Delta \beta$, an analogue of the Cartesian coordinates on the
celestial sphere. In small domains of the sphere such an approximation
is valid, and the calculations may be performed as in Euclidean
geometry. Now the coordinates show how the source coordinates vary from the
Earth's observer viewpoint, while  his moving along the Earth's orbit.

Besides, introduce an auxiliary designation $x_0$, $y_0$ the distance
between the source and the lens at the initial instant in observations
from the barycentre.

The change of Cartesian coordinates will be described by the equations
\begin{equation}
\label{imagemotion1}
x(t)=\mu_{S\lambda} t - p_S \sin (\lambda_E -\lambda_0) +
\left( x_0 +\mu_{\lambda} t -p \sin  (\lambda_E -\lambda_0)\right)
\dfrac{\theta_e^2}{R^2(t)}\\
\end{equation}

\begin{equation}
\label{imagemotion2}
y(t)=\mu_{S\beta} t + p_S \sin \beta_0 \cos (\lambda_E -\lambda_0) +
\left( y_0 +\mu_{\beta} t +p \sin \beta_0 \cos (\lambda_E -\lambda_0)\right) \dfrac{\theta_e^2}{R^2(t)}\\
\end{equation}
where
$$
\label{imagemotion3}
\begin{array}{l}
R^2(t)= x_0^2 +y_0^2 +2 \left( x_0 \mu_{\lambda} +y_0 \mu_{\beta}\right) t + 
\left( \mu_{\lambda}^2 + \mu_{\beta}^2\right) t^2 +\\
2p \left( x_0 \sin (\lambda_E -\lambda_0) -y_0 \sin \beta_0 \cos (\lambda_E -\lambda_0)\right)  + \\
2p \left( \mu_{\lambda} \sin (\lambda_E -\lambda_0) -\mu_{\beta} \sin \beta_0 \cos (\lambda_E -\lambda_0)\right) t + \\
p^2 \left( 1 - \cos^2 \beta_0\cos^2 (\lambda_E -\lambda_0) -2\cos \beta_0 \cos (\lambda_E -\lambda_0)\right)
\end{array}
$$
where the lowest order infinitesimals are retained.

As seen from the equations, (\ref{imagemotion1}, \ref{imagemotion2}) the
image motion depends on several parameters: the proper motion of the source
$\vec \mu_S$, the proper motion of the lens $\vec \mu_D$,  the source
and lens parallaxes $p_S$, $p_D$ and the initial distance between the source
and the lens $(x_0, y_0)$. There  also exists a weak dependence on the initial
declination $\beta_0$, but it does not produce qualitative changes in the
image motion trajectory $S$, and we shall not analyze a dependence of
the trajectory on $\beta_0$.

The distance between the lens and tha lens $(x_0, y_0)$ as well the
source and lens parallax have the same dimensions (they are
dimensionless or measured in angular units, we shall measure these
values in milliseconds of arc).

The proper velocities of the source and lens $\vec \mu_S$ and $\vec
\mu_D$ have dimensions of inverse time (we shall meassure them in
milliseconds of arc per year).

To compare these parameters of different dimensions, it is necessary to
multiply $\vec \mu_S$ and $\vec \mu_D$ by a characteristic time
interval. Such an interval is a year, the interval for which the full
parallax shift of the source and lens is executed.

Analyze the source image trajectory $S$ beginning with the simplest
case of the proper velocities $S$ and $D$ equal to zero $\vec \mu_S=0$
and $\vec \mu_D=0$.

The observer executes only an annual orbital motion, and the solar system
barycentre  rests relative the light source and the lens.

Introduce auxiliary quantities $x_0=\rho \cos \psi$ and $y_0=\rho \sin
\psi$. We shall also assume that the angular distance between the lens
and the  light source is much more than the lens parallax $\Delta n_1
\gg p_D$. This allows the equation for $R^2(t)$ to be simplified.

Write down the simplified equations describing the image motion. To this end
we assume that all velocities in equations (\ref{imagemotion1},
\ref{imagemotion2}, \ref{imagemotion3}), $\vec \mu=0$, $p/\rho\ll 1$. Expand
(\ref{imagemotion1}, \ref{imagemotion2}, \ref{imagemotion3}) as a
series in a small parameter $\dfrac{p}{\rho}$.

Introduce two auxiliary definitions simplifying representation of coordinates
depending on time $A=(p_S-p_D) \cos 2\psi
\dfrac{\theta^2_e}{\rho^2}$, $B=(p_S-p_D) \sin 2\psi
\dfrac{\theta^2_e}{\rho^2}$.

Now the coordinates $x, y$  depending on time (which is determined through the
Earth's right ascension $\lambda_E$) is
\begin{eqnarray}
\label{velocity0}
x -x_S=p_S \sin(\lambda_E-\lambda) -A \sin(\lambda_E-\lambda) +
B \sin \beta \cos(\lambda_E-\lambda) \\
y -y_S=-p_S \sin \beta \cos(\lambda_E-\lambda) -A\sin \beta
\cos(\lambda_E-\lambda) -B \sin (\lambda_E-\lambda)
\end{eqnarray}
where we also introduce auxiliary quantities $x_S=\cos
2\psi\dfrac{\theta^2_e}{\rho}$, $y_S=\sin 2\psi \dfrac{\theta^2_e}{\rho}$.

We shall consider the inequality $A^2 +B^2 \ne p_S^2$ to be valid. Make
a coordinate transformation of the form
\begin{eqnarray}
\label{transform}
\tilde x=(p_S +A) (x -x_S)+ B (y -y_S), \\
\tilde y=-B (x -x_S)+ (A -p_S) (y-y_S).
\end{eqnarray}

\noindent
This transformation incorporates  translation of the origin of coordinates
by the vector $(x_S, y_S)$, turn and dilatation of each of the axes in
the ratio of $\dfrac{p_S +A}{p_S -A}$ and reflection of one of the
axes. The reflection will becomes obvious if one assumes that $A=B=0$, then
the transformations take the form
\begin{eqnarray}
\tilde x=p_S (x -x_S), \\
\tilde y= -p_S (y-y_S).
\end{eqnarray}

The transformation (\ref{transform}) results in equations of the form
$\tilde x=- (p_S^2 -A^2 -B^2) \sin(\lambda_E-\lambda)$,
$\tilde y= (p_S^2 -A^2 -B^2)\sin \beta \cos(\lambda_E-\lambda)$
                                                                           .
If the condition $\Delta = A^2 +B^2 -p_S^2 \neq 0$ is satisfied, then from these equations follows an
equation for a parallax ellipse of the form
\begin{equation}
\tilde x^2 +\dfrac{\tilde y^2}{\sin^2 \beta}=1
\end{equation}

\noindent
in true coordinates ($x, y$) it is a deformed ellipse shifted by the
vector ($x_S, y_S$). In the case of $p_S=0$ the figure is a true ellipse
shifted relative the true position of the source by the vector
$(x_S,y_S)$ and turned  with respect to the ecliptic plane by the
angle $2 \psi$.

It should be noted that a turn of the parallax ellipse is not found in
nonrelativistic astrometry, it is similar to the effect mentioned in the end
of the section "Rigid-Baselined Observations".

\begin{figure}
\epsfxsize=0.7\textwidth
\centerline{\fbox{%
\epsfbox[50 542 400 732]{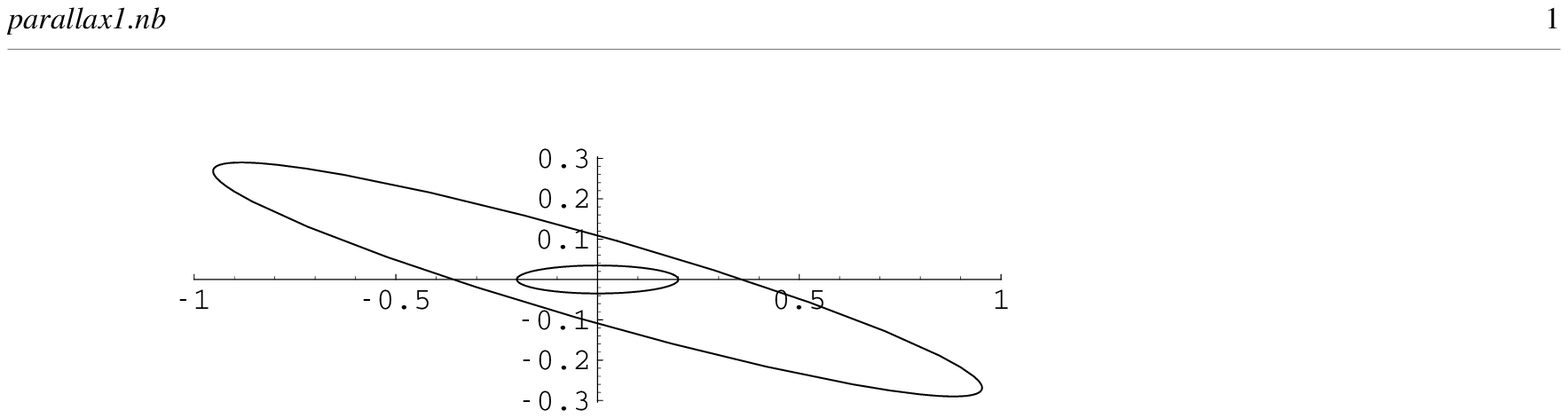}}}
\caption{
The parallax ellipse image of an extragalactic source (a small ellipse
inside a large one) against the background of the parallax motion due to the
weak microlensing effect.
}
\label{fig8}
\end{figure}

Figure (\ref{fig8}) depicts two ellipses: one is a parallax ellipse arising
in the absence of the weak microlensing effect, the other ellipse of larger
size arises in a weak action of a close gravitational lens whose parallax is
much more than the source parallax. For simplicity both ellipses are drawn
coaxially.

In the case of $\Delta=0$ the motion is degenerated into a motion
along a straight line of the form
$$
y-y_S=-\dfrac{p_S+A}{B}( x -x_S)
$$

The parallax motion along a staight line in Euclidean astrometry corresponds
the source position in the ecliptic plane $\beta_0=0$. However in this
case the staight line of the parallax motion is parallel to the axis
$\lambda$. In the case of the parallax motion being due to the weak
microlensing effect there appears a slope of the straight line, with
the slope cofficient depends on the lens parameters
$$
\dfrac{y-y_S}{x -x_S}=\dfrac{p_S}{p_D -p_S} \dfrac{1}{\sin 2 \psi}
\dfrac{\rho^2}{\theta^2_e} +\cot 2 \psi
$$

Thus at a zero proper motion of the lens $D$ and the source $S$, as a
result of an affine transformation of the coordinate system, the image
motion trajectory although the real trajectory possesses a more complex
shape.

Consider in more detail a motion along the parallax curve at different values
of the problem parameters. First of all we shall assume that
$p_S<<p_D$ and introduce a designation $p_g=p_D \dfrac{\theta^2_e}{\rho^2}$.
For simplicity we shall also assume that the Earth's orbit eccentricity is
equal to zero and $\lambda_e = \Omega t$, $\Omega =\dfrac{2 \pi}{1
year}$. Consider the case of $\psi = 0$.  The parallax ellipse equation reads
\begin{equation}
x - x_S = (p_S + p_g) \sin( \lambda_e - \lambda_S) \\
y - y_S = (-p_S + p_g) \sin \beta \cos (\lambda_e - \lambda_S)
\end{equation}
As long as $p_g<p_S$, the visible motion of a celestial source forms the ellipse with
semiaxes $p_S + p_g$ along the $Ox$ axis and $(-p_S + p_g)$ along the
$Oy$ axis. The source motion is clockwise. Compare it with the case of
parallax rigid-baselined measurement.  In case the lens $D$ lies in the
plane formed by the baseline $\vec B$ and the source the parallax increases
(see Fig. 4). The parallax along $Ox$ axis increases as well. Really, the
baseline formed by the Earth, when it holds the positions $x=x_{max}
(y=0)$ and $x=-x_{max}  (y=0)$, in turn forms, together with the
source, a plane wherein lies $D$. An opposite case arises when the
Earth is at the orbit points where the source image occupies positions
$y_{max} (x=0)$ and $-y_{max}(x=0)$. The baseline vector, formed by
these two points of the orbit, is perpendicular to the lens-source
vector. Therewith the rigid-baselined source parallax decreases (see
Fig. 5). Really, the parallax along $Oy$ axis proves to be $(-p_S +
p_D)$ less (in magnitude) than the parallax in the absence of the weak
microlensing effect.

As $p_g$ increases, the parallax along $Ox$ continues to increase, and
along $Oy$ to decrease. In the case of $p_g=p_S$ the source motion
degenerates into a straight line parallel to $Ox$.

A futher increase of $p_g$ results in the parallax along the axis beginning to
grow again, but the source motion will be already retrograde, i.e. clockwise.
This is usually identified with a visible negative parallax.

Calculate a direction of the source motion in the plane $Oxy$. To do this, we
calculate the quantity
$$
J=x \dot y-y \dot x
$$
being an analogue of the $z$ component of the motion moment vector in
mechanics. In our case this is perpendicular to the picture plane. A sign
change in this quantity means an opposite rotation. The quantity $J$ is
$$
J = \dot {\lambda_e} (p^2_S - p^2_g) \sin (\lambda_e - \lambda) \sin \beta
$$
In the absence of the weak microlensing effect $p_g = 0$. The sign of $J$ is
determined by the angular velocity of the Earth orbital motion $\dot
\lambda_e$ and by the sign of $\sin \beta$. A sign change occurs in
the case of $p_g>p_S$. We shall identify the sign change with a visible
negative parallax.  Thus the negative parallax appears under the
condition $p_g>p_S$. It should be noted at once that for most extragallactic
sources one may assume $p_S=0$, and hence the observed parallax will be
negative for them.

Now we consider a more complex case of motion of the image $S$, which
arises at a nonzero proper motion and a nonzero lens parallax. It
should be noted that the case of nonzero lens parallax corresponds to
observation from the barycentre and has been consider previously
\cite{saz98}. The parallax and proper motion of the source will be
assumed zero.

Due to a lens motion there arises an angular motion of the image, as in Fig.
(\ref{fig6}) but along a closed curve since now a motion of the source is
absent. This circular motion is superimposed by an image shift due to the
lens parallax shift (which results in a variable angular distance between the
lens $D$ and the source $S$ with an annual period, see Fig. 10).

Choose the lens parameters as follows. The source declination is
$\beta \approx 30^{\circ}$. The star-lens distance from the solar
system amounts to 200 pc, which corresponds to the lens parallax $p_D=5$
mas. We shall consider that the star-lens mass amounts to $M=2.5
M_{\odot}$, so that Einstein's cone size $\theta_e=10$ mas. The
initial distance between the lens $D$ and the source $S$ is the vector
$\Delta \vec n = (100, 200) mas$.

\begin{figure}
\epsfxsize=0.7\textwidth
\centerline{\fbox{
\epsfbox{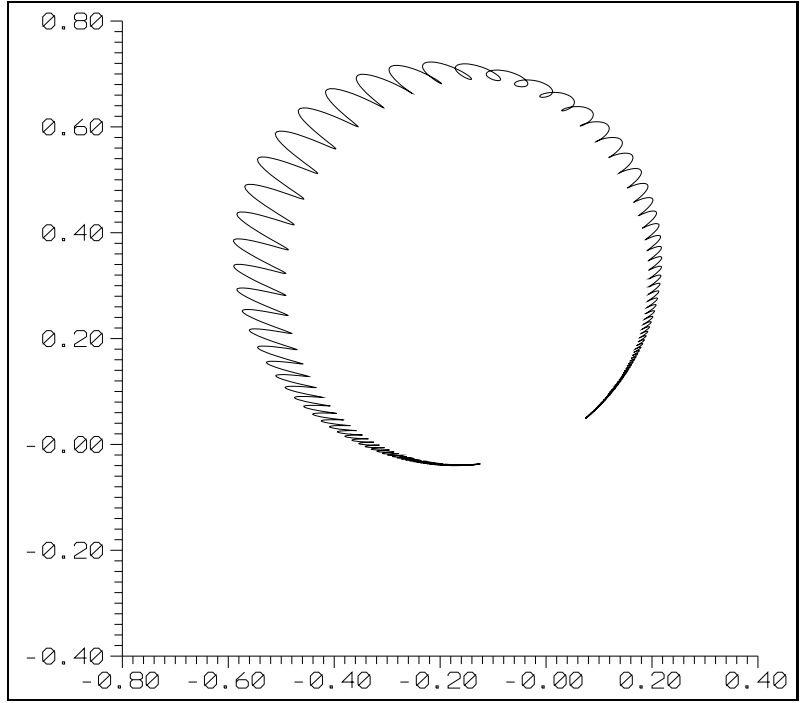}}}
\caption{
The circular motion of the source image is superimposed by distortions
related by the parallax shift of a slow lens.
}
\label{fig9}
\end{figure}

Calculate a trajectory of motion of the image of the source $S$ for
three limiting cases of motion of the lens $D$. The first case is a
slow motion of the lens $|\mu_D|\cdot 1 year <p_D$, in our case we choose parameters of the
proper velocity of the lens $\mu_{\lambda}=1$ mas per year,
$\mu_{\beta}=0.5$ mas per year (see Fig.  \ref{fig9}). In this case
the subtrajectory due to a lens parallax shift (change in the angular
distance between the lens and the source) everywhere fills the large
circle of the motion due to the weak microlensing effect).

\begin{figure}
\epsfxsize=0.7\textwidth
\centerline{\fbox{
\epsfbox{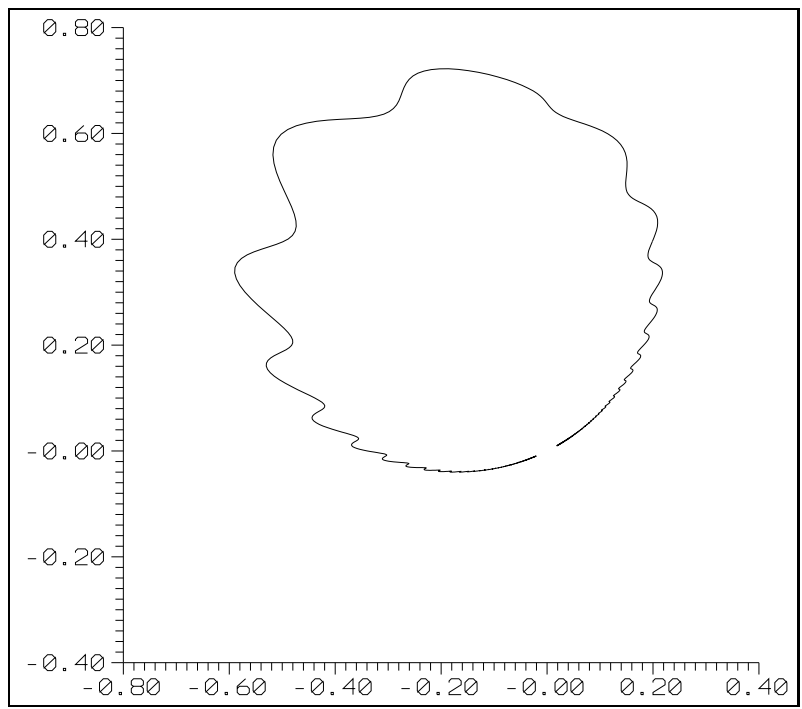}}}
\caption{
The circular motion of the souce image is superimposed by distortions due to
the parallax shift of the lens that covers the angular distance per year
equal to the parallax of the lens itself.
}
\label{fig10}
\end{figure}

In the case of equality of the basic parameters, when the proper motion of
the lens along the right ascension and declination is
$\mu_{\lambda}=10$ mas per year and $\mu_{\beta}=0$ mas per year
respectively, the motion becomes akin to (Fig. \ref{fig10}).

\begin{figure}
\epsfxsize=0.7\textwidth
\centerline{\fbox{
\epsfbox{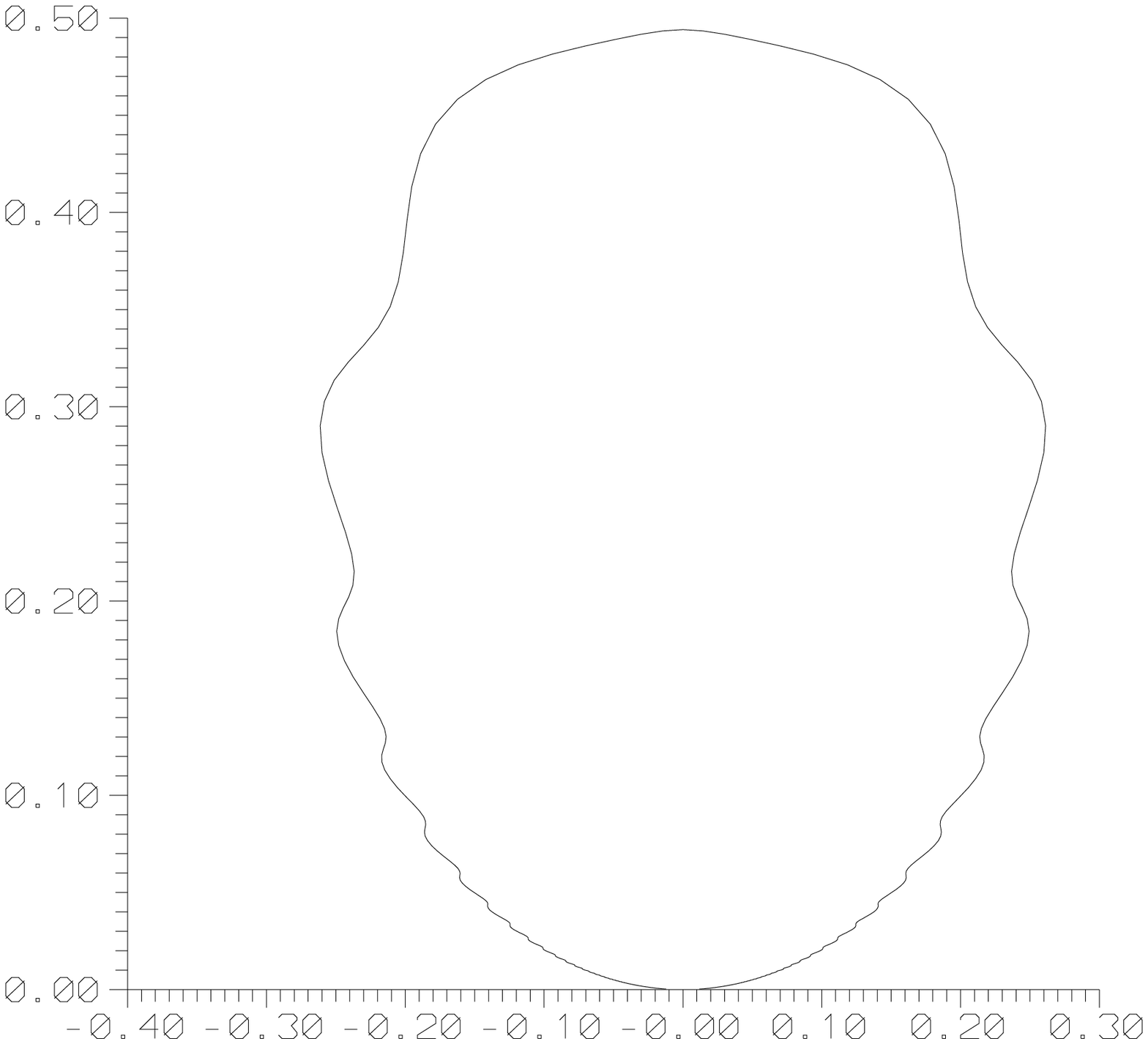}}}
\caption{
The circular motion of the image motion is superimposed by distortions due to
the parallax shift of a fast lens. As seen from the figure, the circle have
converted into a figure similar to the waving circle.
}
\label{fig11}
\end{figure}

Finally, in a fast flight of the lens $\mu_{\lambda}=100$ mas per year
$\mu_{\beta}=0$ the motion is akin to (Fig. \ref{fig11}).

The case considered above, when $p_S=0$ and $\vec \mu_S=0$, is the most natural for
description of the situation arising in the observation of extragalactic
sources. Really, the quasars having considerable red shifts, say $z=0.2$, are
located at distances of the order of $R \sim1$ Gpc. Such a distance
corresponds to the parallax of 1 nas, which cannot be measured in the
neare future.  Random velocities of the quasars amount to several
thousands of km per second. We choose $v\sim 3000$ km per second. Such
velocities of the transversal motion correspond to the proper angular
velocities, when the object is at a distance of 1 Gpc, of the order
$\mu < 1 {\mu as \over year}$.  This value may be also neglected.

\begin{figure}
\epsfxsize=0.7\textwidth
\centerline{\fbox{
\epsfbox{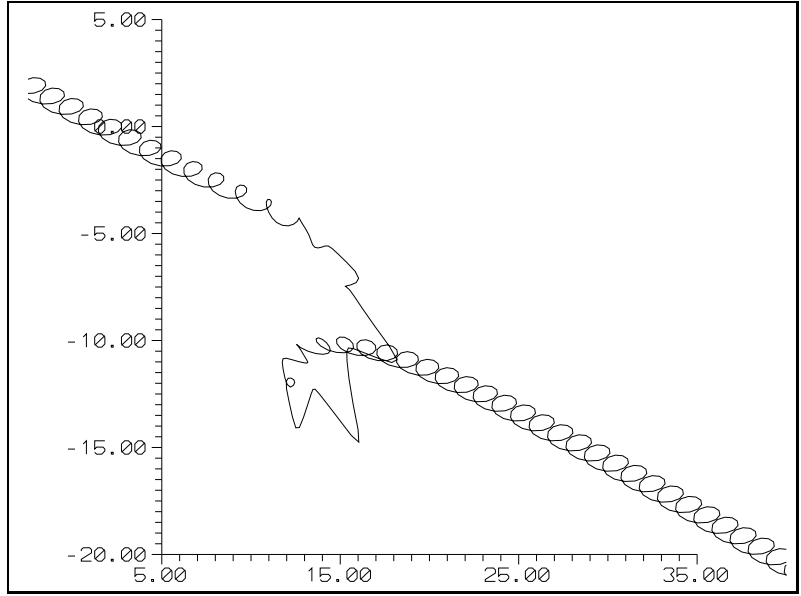}}}
\caption{
The source image motion is superimposed by distortions related to a circular
motion due to the weak microlensing effect, lens and source parallax shifts.
}
\label{fig12}
\end{figure}

However, while discussing the weak microlensing effect, when the lens and the
source are in our Galaxy (although the lens is closer than the source), the
parallax of the source and its proper motion cannot be neglected. In this
case there arises the most complex visible motion of $S$ (see Fig.
\ref{fig12}).

Fig.\ref{fig12} depicts a model with the object parameters as follows.
The source declination is equal to $30^{\circ}$; the proper motion of
the lens $\mu_{\lambda}=8$ mas per year, $\mu_{\beta}=3$ mas per year.
The proper motion of the lens $\mu_{\lambda}=1$ mas per year,
$\mu_{\beta}=-0.5$ mas per year. The values of the parallax shift for
the source and the lens are equal $p_S=1$ mas and $p_D=3$ mas
respectively.  Einstein's cone size $\theta_e=10$ mas, which
corresponds to an object with the mass about four Solar masses. The
time interval during which the modelling has been conducted is equal to
40 years.

The visible motion of the source appears to be especially involved in
case the minimum angular distance between the lens $D$ and the light
source $S$ is comparable with the parallax of the lens itself. In the
case of a slow motion of the lens there may arise a situation of
"transferring" the image $S$ resulting in large visible displacements
of $S$ on the sky.  Therewith the motion of the image approaches a
chaotic one.  Although the case of $\Delta n \sim p_D$ is unlikely, the
chaotic motion of the source $S$ will mean realization of just such a
case.

\section*{Conclusion}

Two important conclusions can be made. The first one lies in a possible
appearance of rather a complex visible motion of the extragalactic
sources, which are ICRF basic sources, under the action of the weak
microlensing effect. This motion may exceed considerably the proper
notion of a part of the sources, with the motion due to the weak
microlensing effect being composed of two motions -- one having a
characteristic angular quantity $\delta \varphi \sim
\theta^2_e/\theta_i$ (here $\theta_i$ is the angular distance between
the lens and the source) and a characteristic time of tens and hundreds
of years, the other having a period $\sim 1$ year related to the the
lens parallax motion and the quantity of angular shift $\sim p_D
(\dfrac{\theta_e}{\theta_i})^2$. Although the second shift is less than
the first one for most events, it may amount to a good part of the
total angular deviation $\delta \varphi$.

In paper (\cite{saz98}) an effect of a weak microlensing on ICRF
sources was considered. 8 sources were found to belong to the list of
607 ICRF sources whose angular shift due to the weak microlensing
effect was maximal. Parallax changes can be simulated for these
sources. On conducting the simplest simulation, it has been found that
there appears a retrograde parallax motion due to the wreak
microlensing effect. The value of this parallax averages 2 $\mu as$
varying from 50 nas to 5 $\mu as$. Since the investigated sources are
assumed to be extragalactic ones, all parallaxes are negative.

Hence the second conclusion lies in the motion being retrograde in most
cases, which corresponds negative parallaxes. The values of parallaxes
due to the weak microlensing effect is of the order of hundreds of
nanoseconds and microseconds of arc.

\section*{Acknowledgement}

The authors are grateful to S. Kopeikin, A. Kuzmin, K. Kuimov for
valuable remarks and comment. This work has been done under support of
"Cosmion" center, Russian Fund of Basic Research (grants NN
98-05-64797, 00-02-16350), as well as "Russian Universities" Programme
(grants N 2-5547, N 9900777).


\newpage


\begin{thebibliography}{99}

\bibitem[1]{b1}
Alcock C. et al., 1993, Nature, 365, 621

\bibitem[2]{z7}
Allen C.W. ed., 1973, Astrophysical Quantities. Univ. of London,
The Athlone Press

\bibitem[3]{and81}
Andreyanov V.V., Kardashev N.S., 1981 Kosmicheskie issledovaniya 19, 763.
(in Russian)

\bibitem[4]{and86}
Andreyanov V.V et al., 1986, Astronom. Zh.,
63, 850 (in Russian)

\bibitem[5]{b2}
Aubourg E. et al., 1993, Nature, 365, 623

\bibitem[6]{bli89}
Bliokh, Minakov, 1989, Gravitational lenses, Kiev, Naukova dumka (in
Russian)

\bibitem[7]{das65}
Dashevsky, Zeldovich, 1965, Astron. Zh., 41. 1071. (in Russian)


\bibitem[8]{eic}
Eichorn H., Astronomy of stars positions.
Frederick Ungar Pub. Co., New York.


\bibitem[9]{eub98}
Ma C et al., Astron. J. 116, 516, 1998.

\bibitem[10]{fei99}
Gontier A.-M., Feissel M., Essaifi N., Jean-Alexis D., Paris
Observatory Analysis Center OPAR on activities, Jan98 - Mar99.

\bibitem[11]{gre}
Green R. M. Spherical Astronomy. Cambridge Univ.Press, 1985.

\bibitem[12]{hip}
The Hipparcos and Tycho Catalogues. Vol.1, Introduction and Guide
to the Data. M.A.C. Perryman. ESA Publ. Div., c/o ESTEC, Noordwijk,
The Netherlands, June 1997.

\bibitem[13]{hog}
Hog~E., Novikov, Polnarev, 1994,  Nordita Preprint,
Macho Photometry and Astrometry, Nordita -- 94/26 A

\bibitem[14]{iers96}
IERS, 21, 1996, International Earth Rotation Service Annual report,
Observatoire de Paris

\bibitem[15]{z1}
IERS, 1995, 1994 International Earth Rotation Service Annual
report, Observatoire de Paris

\bibitem[20]{z2}
Jacobs C.S., Sovers O.J., Williams J.G., Standish E.M., 1993,
Advances in Space Research, 13, No. 11, 161

\bibitem[21]{z5}
Jenkner H., Lasker B.M., Struch C.R., McLean B.J., Shara M.M,
Russel J.L., 1990,  AJ, 99, 2082

\bibitem[22]{z8}
Kaplan S.A., Pikelner S.B., 1979,
Physics of Interstellar Medium, Moscow, Nauka (in Russian)


\bibitem[23]{kar86}
Kardashev N.S., 1986, Astron. Zh., 63. 845. (in Russian)

\bibitem[24]{kop}
Kopeikin S., Scheffer Phys. Rew, D, 60, N124002, 1999
gr-qc 9902030.

\bibitem[25]{kop00}
Kopeikin S., Gwinn C., Sub - Microarcsecond Astrometry and New
Horizons in Relativistic Gravitational Physics. Proc. IAU Coll., 180.

\bibitem[26]{z3}
Lasker B.M., Struch C.R., McLean B.J., Russel J.L., Jenkner H.,
Shara M.M., 1990,  AJ, 99, 2019


\bibitem[27]{mar}
Murrey C.A., Vectorial astrometry, Royal Greenwich Observatory,
Herstmonceux Castle, East Sussex,  Adam Hilder, Bristol, 1983.

\bibitem[28]{iers}
McCarthy D.D., 1996, ed., IERS Conventions. IERS Technical Note
21, Observatoire de Paris

\bibitem[29]{pac86}
Paczinsky B., 1986, ApJ, 304, 1

\bibitem[30]{z6}
Perryman M.A.C. et al., 1997, A\& A, 323, L49


\bibitem[31]{gai98}
Project GAIA. http://astro.estec.esa.nl/SA-general/Project/GAIA;

\bibitem[32]{sim98}
Project SIM. http://sim.jpl.nasa.gov/

\bibitem[33]{dar98}
Project DARWIN. http://ast.star.rl.ac.uk/darwin;

\bibitem[34]{faim}
Project FAME. http://aa.usno.navy.mil/fame/

\bibitem[35]{diva}
Project DIVA. http://www.aip.de/groups/DIVA/

\bibitem[36]{ref64}
Refsdal, 1964, Monthly Not. Roy. Astron. Soc., 128, 295.

\bibitem[37]{z4}
Russel J.L., Lasker B.M., McLean B.J., Struch C.R., Jenkner H., 1990,
AJ, 99, 2059


\bibitem[38]{saz96}
Sazhin M.V., 1996, Pis`ma v Astron. Zh., 22, 643 (in Russian)

\bibitem[39]{saz98}
Sazhin M.V., Zharov, A.V.Volynkin, Kalinina T.A., 1998, Monthly Not. Roy.
Astron. Soc., 300, 287

\bibitem[40]{sch}
Schneider P., Ehlers J., Falco E.E. Gravitational Lenses.
Berlin, New York, Springer Verlag,1992

\bibitem[41]{b3}
Udalski A., Szymanski M., Kaluzny J., et al., 1994, ApJ Lett., 426, L69

\bibitem[42]{zak97}
Zakharov A.F., Gravitational Lenses and Microlenses. Moscow.
Janus-K Publ., 1997. (in Russian)

\bibitem[43]{zak98}
Zakharov A.F., Sazhin M.V. 1998, Physics - Uspekhi, 41, 945.

\bibitem[44]{zel64}
Zeldovich, 1964, Astron. Zh., 41, 19.(in Russian)

\bibitem[45]{zda95}
Zhdanov I.I, et al., 1995, Astron. and Astrophys., 299, 321.


\end{thebibliography}
\end{document}